\documentclass[twocolumn,prl,amssymb,amsmath,natbib,showpacs,floatfix]{revtex4}
\usepackage{epsfig}
\usepackage{graphicx}
\usepackage{bbold}
\usepackage{mathptmx}
\usepackage{bm}

\begin{document}

\title{Multiple Components in Narrow Planetary Rings}%
\author{L. Benet}%
\email{benet@fis.unam.mx}%
\affiliation{%
  Instituto de Ciencias F\'{\i}sicas,
  Universidad Nacional Aut\'onoma de M\'exico (UNAM),
  Cuernavaca, Mor., M\'exico }%
\author{O. Merlo}%
\email{merlo@fis.unam.mx}%
\affiliation{%
  Instituto de Ciencias F\'{\i}sicas,
  Universidad Nacional Aut\'onoma de M\'exico (UNAM),
  Cuernavaca, Mor., M\'exico }%

\date{\today}

\begin{abstract}
  The phase-space volume of regions of regular or trapped motion, for
  bounded or scattering systems with two degrees of freedom
  respectively, displays universal properties. In particular, drastic
  reductions in the volume (gaps) are observed at specific values of a control
  parameter. Using the stability resonances we show that they, and not
  the mean-motion resonances, account for the position of these
  gaps. For more degrees of freedom, exciting these resonances divides
  the regions of trapped motion. For planetary rings, we demonstrate
  that this mechanism yields rings with multiple components.
\end{abstract}


\maketitle

The Cassini mission is providing unprecedented amounts of data on the
structure of Saturn rings~\cite{Cassini}. As usual, surprises and new
questions emerge which call for theoretical understanding. Beyond the
intrinsic interest on planetary rings~\cite{rings}, these are now the
only examples of flat astrophysical systems for which such detailed
data is available. This makes their analysis
paradigmatic~\cite{rings}, which serves as analog for other systems
like galaxies and planet-forming disks. A first order understanding of
the dynamics and processes in rings has been obtained, but important
questions remain unanswered~\cite{rings}. In particular, narrow
planetary rings are eccentric and display multiple components (or
strands), kinks, clumps and arcs (or patches); the F ring of Saturn is
a magnificent case~\cite{Cassini,rings}. Therefore, these narrow rings
pose interesting questions in dynamical astronomy and nonlinear
dynamics related to their stability, confinement and structural
properties.

Recently it became clear that the phase-space volume associated to
regular orbits for bounded systems with two degrees of freedom (DOF)
displays universal properties as a function of a control
parameter~\cite{Contopoulos,Dvorak05}. For scattering systems such
quantity is related to the trapped orbits. In particular, there are
locations where a drastic reduction of the phase-space volume occupied
by the regular or trapped orbits is observed, i.e., gaps. Here we show
that these gaps are related to the occurrence of stability resonances
(SR), and not to mean-motion resonances (MMR). The MMR correspond to
locations where the period of the particle is a rational of the
orbital period of a satellite or planet. Examples include Kirkwood
gaps in the asteroid main belt~\cite{asteroids,histogram} and the
Cassini division in Saturn rings~\cite{MurrayDermott99}, where very
few particles are observed. Other cases of MMR exist where particles
accumulate locally, as the Jupiter Trojans or the Hilda
group~\cite{resonances_revs}. The SR are defined by a resonant
condition in the (linear) stability exponents of a central stable
periodic orbit. We show in this Letter that nonlinear and
higher-dimensional effects related to the SR yield rings with two or
more components, or strands, as those observed in the F
ring~\cite{Cassini,rings}. We use the scattering approach to narrow
planetary rings~\cite{Merlo07} and a disk rotating on a Kepler orbit
as example~\cite{Meyer95}. Our results can be used in other cases,
ranging from particle accelerators~\cite{aceleradores} to galactic
dynamics~\cite{galaxias}, where resonances and transport properties
are important~\cite{Zaslavsky}.

The scattering approach to narrow rings~\cite{Merlo07} considers the
scattering dynamics of a (ring) particle in a planar system with some
intrinsic rotation (external force). In the planetary case, this
rotation comes from the orbital motion of one or several
satellites. This rotation creates generically phase-space regions of
{\it dynamically trapped motion}, in what otherwise is dominated by
unbounded trajectories; ``generic'' implies that it holds for a wide
class of rotating potentials~\cite{Benet00}, including gravitational
interactions~\cite{Benet01,Merlo07}. For a uniformly circular
rotation, the system has two degrees of freedom (DOF) and a constant
of motion, the Jacobi integral. For these scattering systems the
organizing periodic orbits generically appear in pairs through
saddle-center bifurcations, one is stable and the other unstable. The
manifolds of the unstable orbit intersect, isolating a region where
trapped motion is of nonzero measure if stable orbits exist. By
changing the Jacobi integral, the central {\it linearly} stable
periodic orbit becomes unstable and a period doubling cascade sets
in. This scenario turns eventually the horseshoe (invariant set) into
a hyperbolic one, thus destroying any region of trapped
motion~\cite{Rue94}. Then, in the Jacobi constant space, the regions
of trapped motion are bounded. Particles with initial conditions
outside these regions escape rapidly along scattering trajectories;
those inside remain dynamically trapped. Consider now an ensemble of
initial conditions of independent particles distributed over the
extended phase space, which contains entirely one region of trapped
motion. Then, due to the intrinsic rotation, the pattern formed by
projecting the trapped particles into the $X-Y$ plane at a given time
forms a ring~\cite{Benet00}. The ring is typically narrow, sharp-edged
and noncircular; this scenario is generic~\cite{Benet00}. For more
than two DOF, e.g., a nonuniform rotation on a Kepler elliptic orbit,
the ring displays further structure~\cite{Merlo07}.

The simplest example to illustrate the occurrence of rings is a planar
scattering billiard: a disk on a Kepler orbit~\cite{Meyer95}. This
system consists of a point particle moving freely unless it collides
with an impenetrable disk of radius $d$, which orbits around the
origin on a Kepler orbit of semi-major axis $R$ ($d<R$). Collisions
with the disk are treated as usual~\cite{Meyer95}; no collisions lead
to escape. For the disk on a circular orbit, the simple periodic
orbits are the radial collision orbits. They and their linear
stability are given by~\cite{Merlo07}%
\begin{equation}
\label{eq1}
J =  2\omega_d^2 (R-d)^2 
(1+\Delta\phi\tan\theta) \cos^2\theta(\Delta\phi)^{-2},
\end{equation}
\begin{equation}
  \label{eq2}
  {\rm Tr} D{\cal P}_J  =  2 +
  \big[(\Delta\phi)^2(1-\tan^2\theta)-4(1+\Delta\phi\tan\theta)\big]R/d.
\end{equation}
Equation~(\ref{eq1}) defines the value of the Jacobi integral $J$ for
the radial collision orbits in terms of $\theta$, the outgoing angle
of the particle's velocity after a collision with the disk. Here,
$\Delta\phi=(2n-1)\pi+2\theta$ is the angular displacement of the disk
between consecutive radial collisions and $n=0,1,2,\dots$ is the
number of full turns completed by the disk before the next
collision. The period of the disk's orbit is $T_d=2\pi$
($\omega_d=1$). Equation~(\ref{eq2}) provides the trace of the
linearized matrix $D{\cal P}_J$ around the radial collision
orbits. Being a two DOF system, periodic orbits are {\it linearly}
stable iff $|{\rm Tr\,} D{\cal P}_J|\le2$.

To understand the phase-space properties that define the dynamics, in
particular, for many DOF, we consider a relative measure of the
phase-space volume occupied by the regions of trapped
motion~\cite{Contopoulos,Dvorak05}. This quantity is a function of
some control parameter and tunes the horseshoe
development~\cite{Rue94}. For the disk on a circular orbit a good
choice is $J$. However, this quantity is not conserved for nonzero
eccentricity $\varepsilon$, thus being useless for more than two
DOF. A convenient quantity is the average time between consecutive
collisions with the disk, $\langle\Delta t\rangle$. The average is
defined, for a given initial condition (ring particle), over the
successive consecutive collisions times; in addition, we consider an
ensemble of them. In our numerical calculations we considered an orbit
to be trapped if it displays more than $20000$ collisions with the
disk; the next $200000$ reflections were used to improve the
statistics.

\begin{figure}
  \centerline{\includegraphics[angle=270,width=9.5cm]{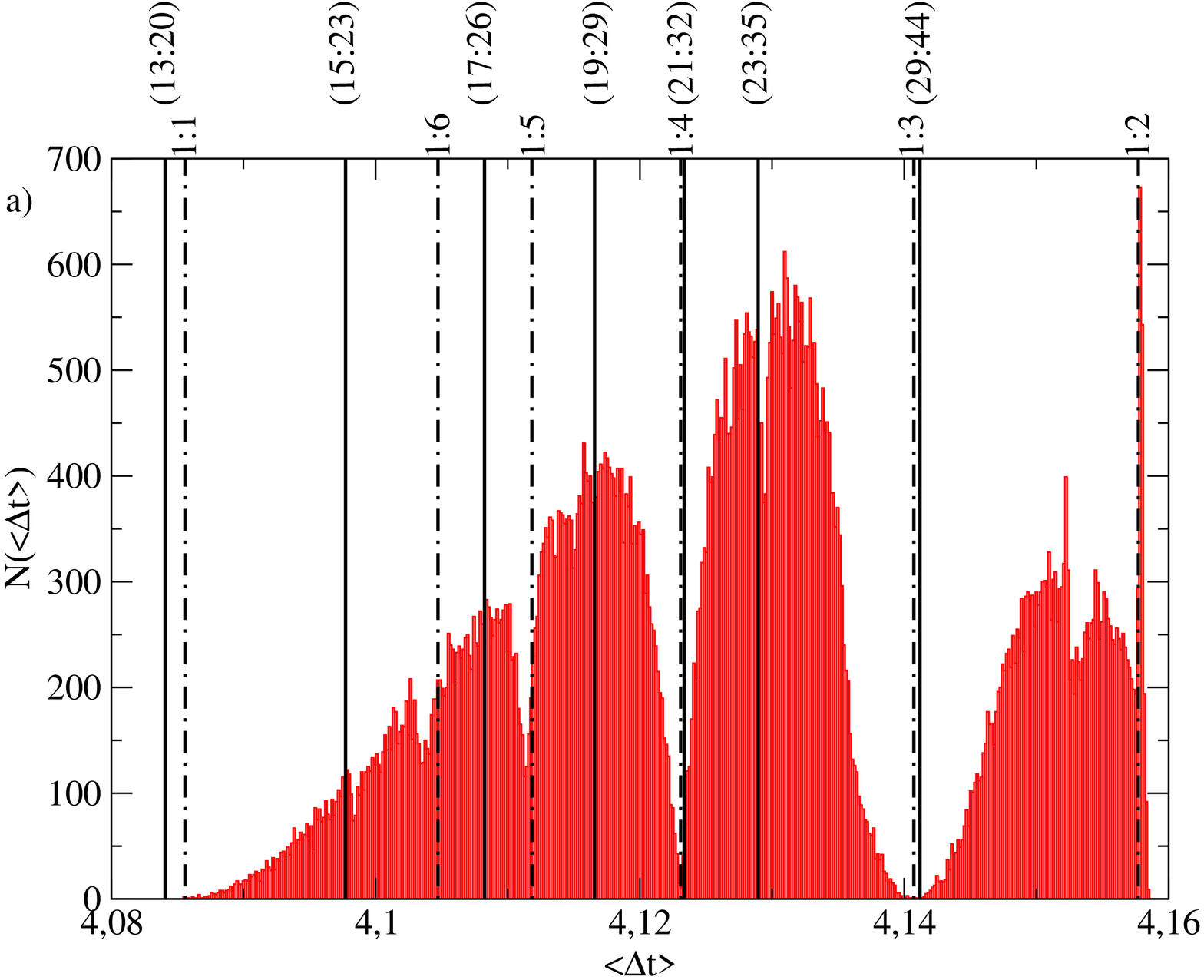}}
  \vskip -10pt
  \centerline{\includegraphics[angle=270,width=9.5cm]{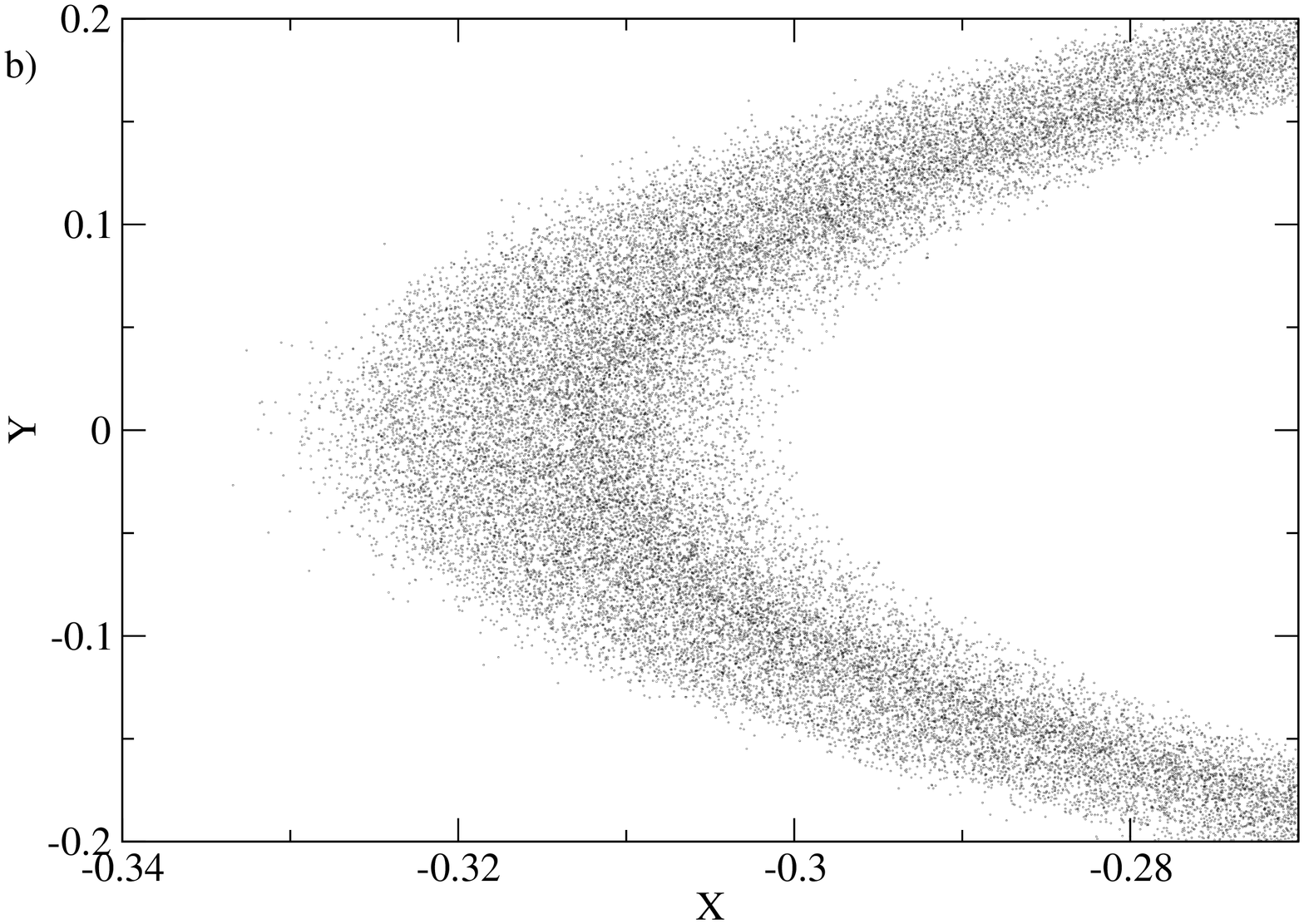}}
  \caption{\label{fig1}%
    (Color online) (a)~Histogram of $\langle\Delta t\rangle$ showing
    the relative phase-space volume of the region of trapped motion
    for the scattering billiard on a circular orbit. The continuous
    lines represent the position of the lowest MMR, indicated in
    parentheses, that occur in the interval of $\langle \Delta t
    \rangle$; dash-dotted lines correspond to the lower-order
    SR. (b)~Detail of the ring corresponding to this region of trapped
    motion.}
\end{figure}

In Fig.~\ref{fig1}(a) we show the histogram of $\langle\Delta
t\rangle$ for the $n=0$ trapped region; Fig.~\ref{fig1}(b) shows a
detail of the corresponding ring~\cite{Merlo07}. We note that
Fig~\ref{fig1}(a) displays some structure, namely, some rather well
localized positions where there is essentially no trapped motion in
phase space or, at least, a drastic reduction of its size. This
structure is similar to those in Ref.~\cite{Contopoulos}, and is
universal~\cite{Contopoulos,Dvorak05}. Then, universality extends to
scattering systems in terms of the phase-space volume of the regions
of trapped motion. We note that the structure of Fig.~\ref{fig1}(a)
also resembles the structure of the population histogram of asteroids
in terms of their semimajor axis~\cite{histogram}, which uncovers
the Kirkwood gaps and the role of MMR.

Guided by the similarity with the asteroids, we ask whether the gaps
in Fig.~\ref{fig1}(a) are related to MMR. The question can be answered
analytically for the radial collision periodic orbits which are the
organizers of the dynamics. Their collision time is $t_{\rm
  col}=\Delta\phi/\omega_d$, hence the MMR condition is $t_{\rm
  col}/T_d = p/q$, with $p$ and $q$ incommensurate integers. In
Fig.~\ref{fig1}(a), we have indicated the location of the {\it lowest}
mean-motion resonances in the $\langle\Delta t\rangle$ interval of
interest; the actual resonances are indicated in parentheses. We have
also included the 29:44 resonance because of its proximity to the
position of the main gap. While these resonances are definitely not
low-order resonances, we note that some may be very close to the
gaps, while others are not. Therefore, low-order mean-motion resonances
are not related to the reduction of the phase-space volume (gaps) of
the regions of trapped motion.

\begin{figure*}
  \centerline{%
    \includegraphics[angle=270,width=6.5cm]{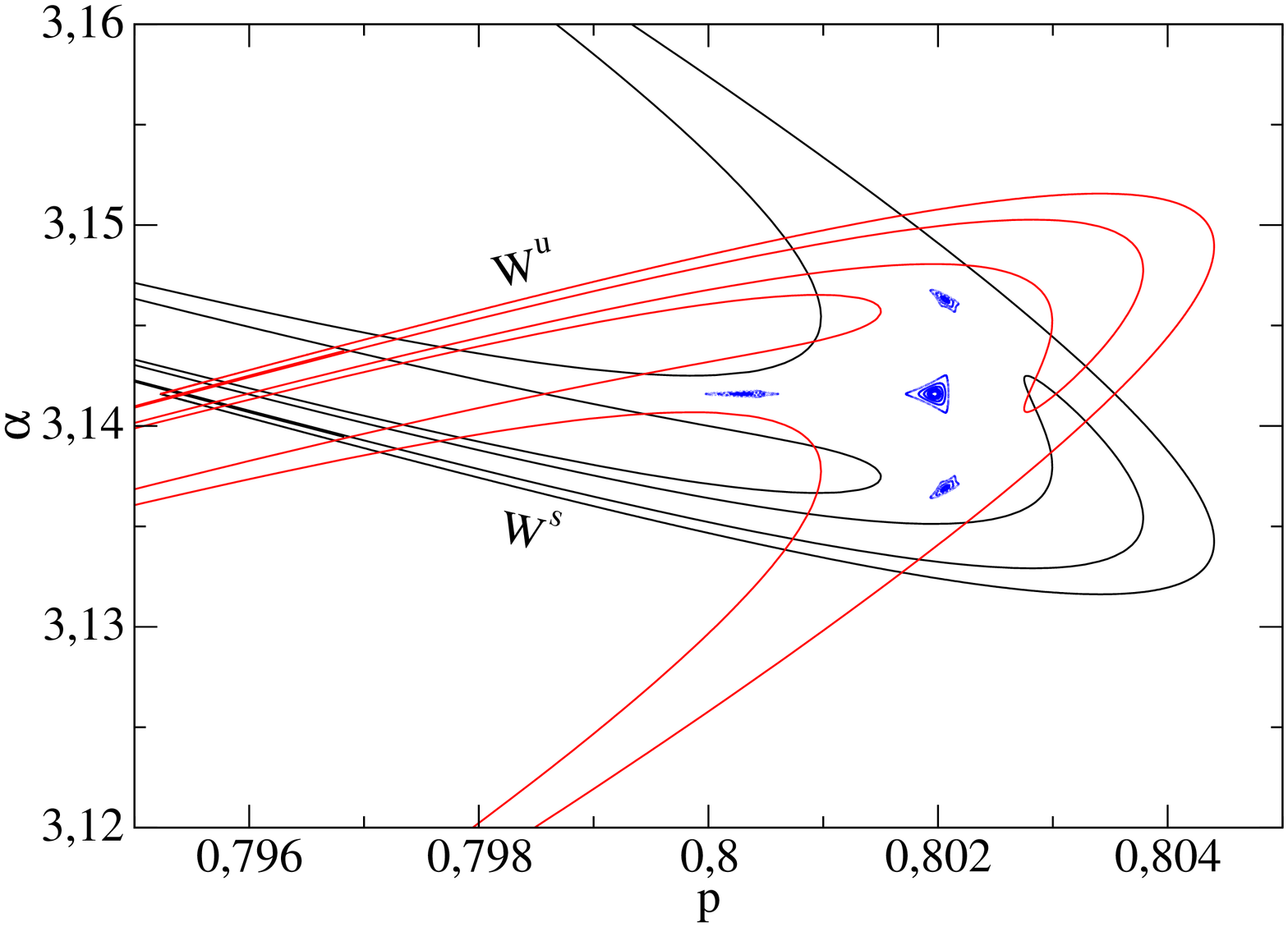}
    \includegraphics[angle=270,width=6.5cm]{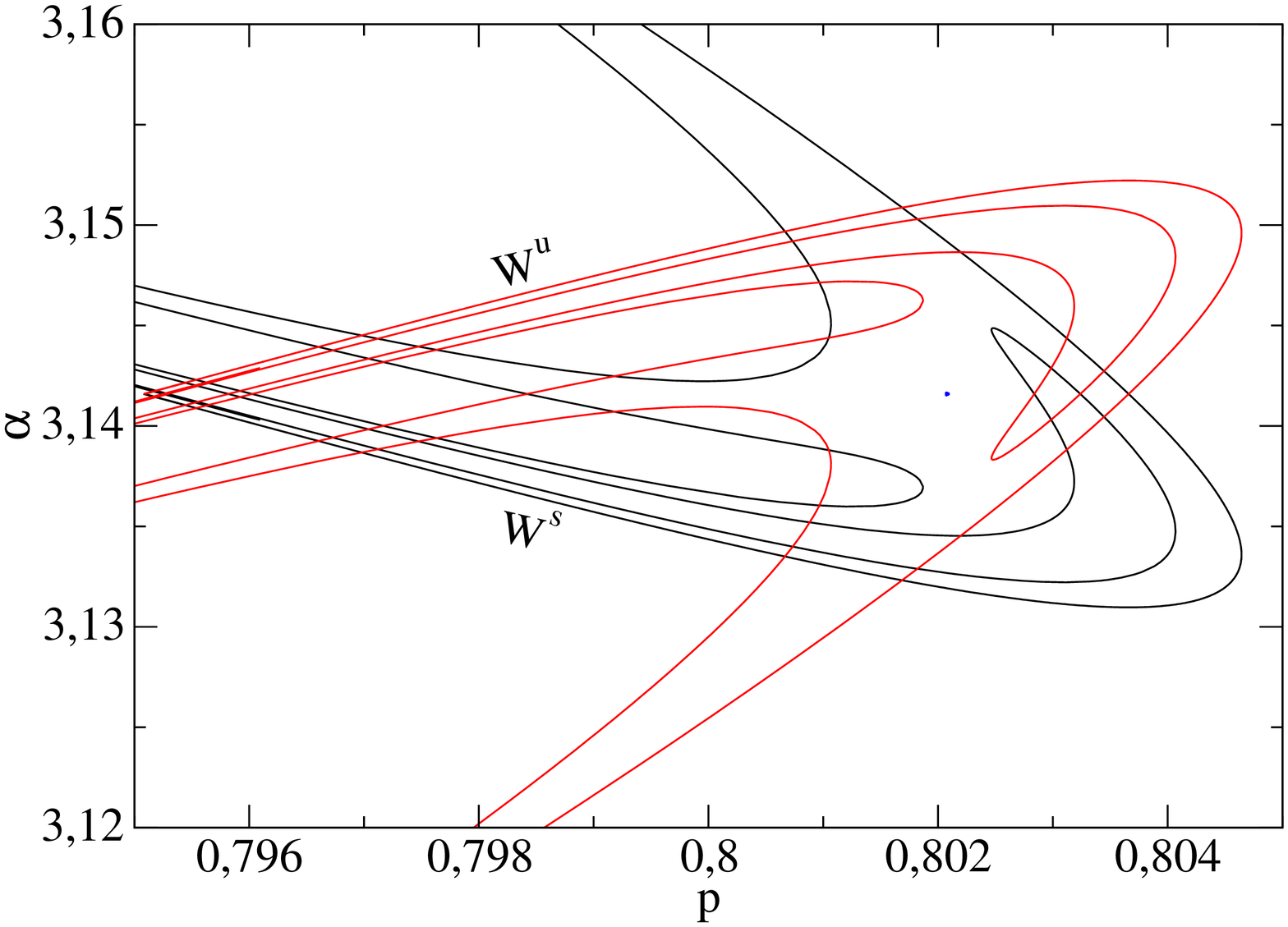}
    \includegraphics[angle=270,width=6.5cm]{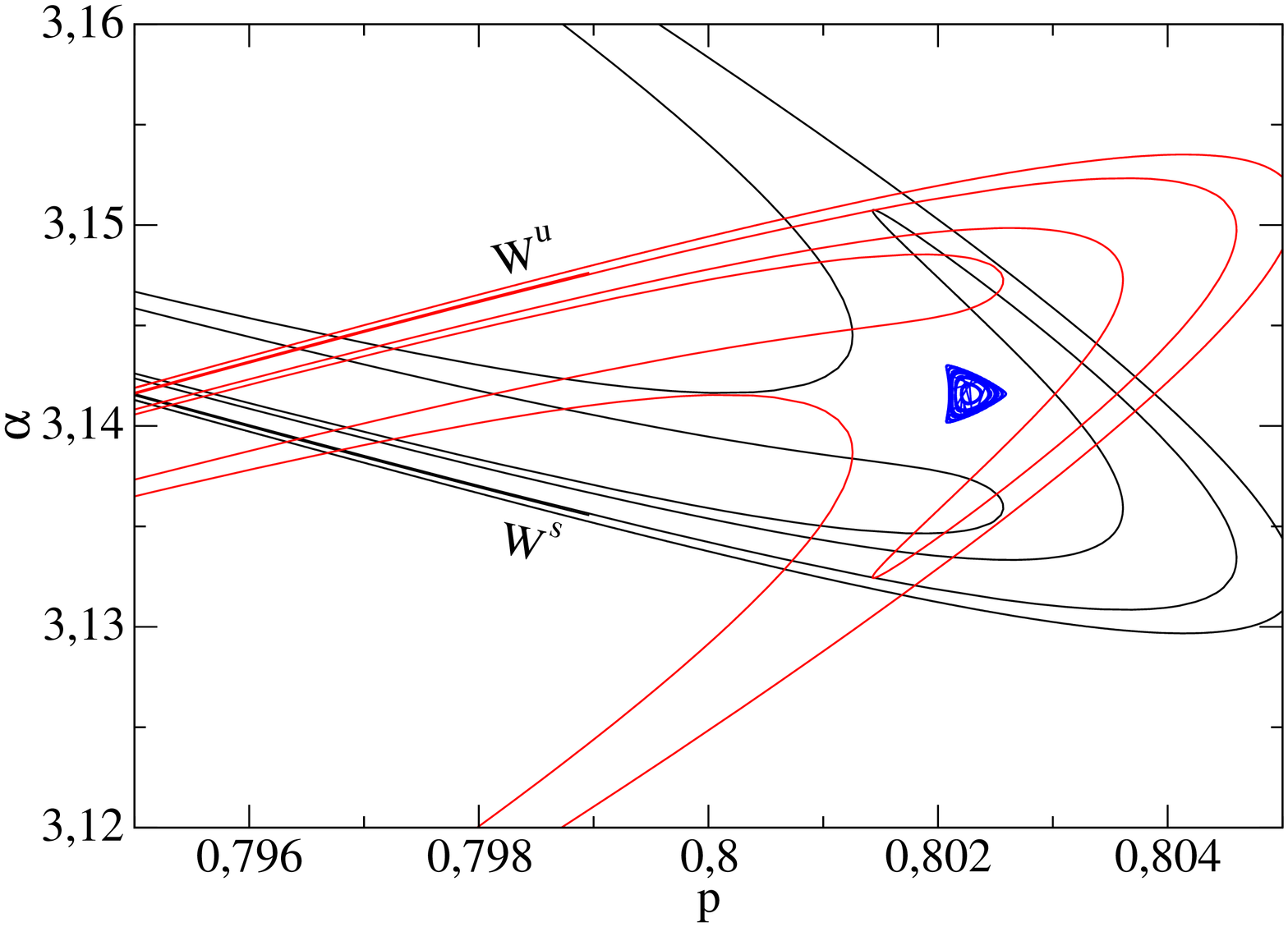}}%
  \caption{\label{fig2}%
    (Color online) Surfaces of section for the scattering billiard on
    a circular orbit for decreasing values of $J$. The sequence
    illustrates the reduction of the phase-space volume of the trapped
    region due to the 1:3 SR corresponding in Fig.~\ref{fig1}(a), from
    left to right, to $\langle \Delta t \rangle \approx 4.1382$,
    $\langle \Delta t \rangle \approx 4.1409$, and $\langle \Delta t
    \rangle \approx 4.1445$. The black/red curves are the
    stable/unstable manifolds of the unstable periodic orbit. The blue
    regions correspond to regions that represent more than 90\% of the
    phase-space volume occupied by the trapped orbits.}
\end{figure*}

To understand the gaps of Fig.~\ref{fig1}(a), we consider the
phase-space structure defining a surface of section at the collisions
with the disk in a rotating frame (see~\cite{Merlo07}). Thus, the
average return time to the surface of section is precisely
$\langle\Delta t\rangle$. Figures~\ref{fig2} display the surfaces of
section for three values of the Jacobi integral around the location of
the main gap in Fig.~\ref{fig1}(a). Satellite period-three islands
squeeze the central stable fixed point, reducing efficiently the
region of trapped motion. At some value of $J$ there is a bifurcation
with the central point being unstable. After this value, the whole
structure reappears with some symmetric inversion. This phenomenon is
intrinsically nonlinear and is universal~\cite{Arnold}. Note that
the existence of the period-three islands in the surface of section
is not equivalent to a 1:3 MMR. The period-three fixed points
involve different collision times. Adding these individual times
matches a multiple of the collision time of the central fixed point,
which in general is not a rational of $T_d$.

The gaps in Fig.~\ref{fig1}(a) are thus related to a reduction of the
phase-space volume occupied by trapped orbits due to satellite islands
(see also~\cite{Simo}). In ~\cite{Arnold} it is shown that the passage
through such 1:3 resonances (Fig.~\ref{fig2}) leads universally to
instabilities, while resonances of order higher than 4 do not induce
instability; for the resonance of order 4 the behavior depends upon
which contribution (resonant or nonresonant) dominates the normal
form~\cite{Arnold,Simo}. These results follow from the structure of
the resonant normal form, i.e., the relevant (nonlinear) contributions
to the stability analysis of a central linearly stable periodic
orbit. We thus consider whether these resonances are related to the
structure of Fig.~\ref{fig1}(a). We write the eigenvalues of the
linearized map $D{\cal P}_J$ around the stable radial collision orbit
as $\lambda_\pm=e^{\pm i\alpha}$. We are interested in values of
$\alpha$ that are rational multiples of $2\pi$, i.e.,
$\alpha/(2\pi)=p/q$ with $p$ and $q$ incommensurate integers. The
eigenvalues $\lambda_\pm$ are related to each other by complex
conjugation; by consequence, the $p:q$ resonance is related to the
$q-p:q$ resonance. Since these resonances involve the stability
exponents, we refer to them as stability resonances.

From the definition of the purely imaginary eigenvalues $\lambda_\pm$,
$|{\rm Tr\,} D{\cal P}_J|\le 2$, we have $\cos\alpha={\rm Tr\,} D{\cal
  P}_J/2$. This can be written in terms of the collision time $t_{\rm
  col} = \Delta\phi/\omega_d$ using Eq.~(\ref{eq2}). In
Fig.~\ref{fig1}(a), we have indicated the lower-order SR as
dash-dotted lines. The correspondence with the gaps is
astonishing. Therefore, we attribute the gaps in Fig.~\ref{fig1}(a) to
crossing a SR.

\begin{figure}
  \centerline{\includegraphics[angle=270,width=9.5cm]{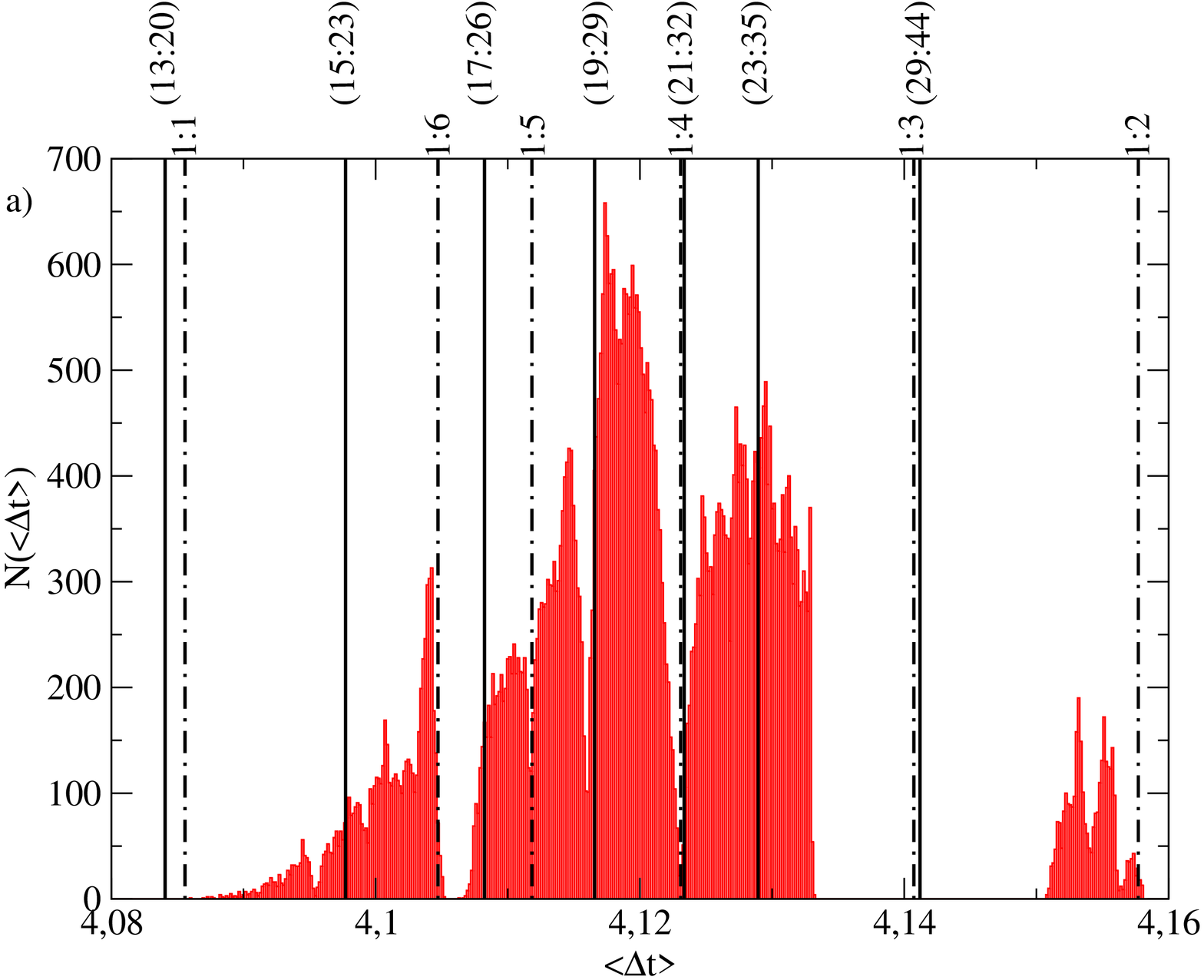}}
  \vskip -10pt
  \centerline{\includegraphics[angle=270,width=9.5cm]{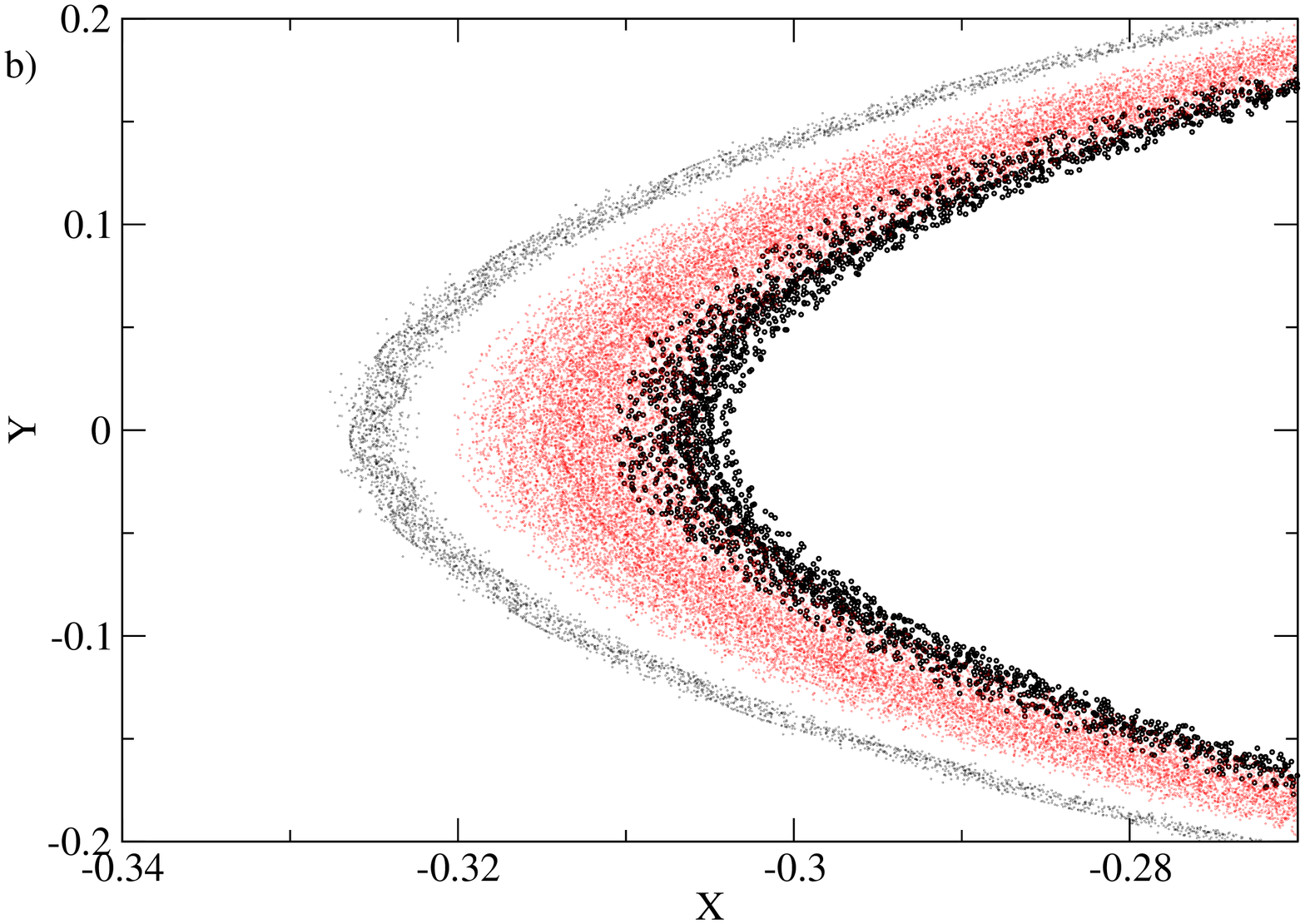}}
  \caption{\label{fig3}%
    (Color online) Same as Fig.~\ref{fig1} for the billiard on an
    elliptic Kepler orbit with eccentricity $\varepsilon=0.0001$. The
    resonances indicated correspond to $\varepsilon=0$. In comparison
    to the circular case, the gap of the 1:3 SR is wider. This causes
    a true opening in the corresponding ring, thus forming a two-component
    ring. The black inner component of the ring corresponds to the
    region on the left of the 1:6 SR; the gray (red) is associated to
    the region between the 1:6 and 1:3 resonances. Note that the 1:6
    resonance gap is not wide enough to have a projection with another
    (distinct) strand.}
\end{figure}

The SR, and, in particular, the 1:3, have interesting consequences in
the context of rings when we consider a small nonvanishing
eccentricity of the disk's orbit. Note that, for nonzero
$\varepsilon$, the system is explicitly time dependent, has
two-and-half DOF and cannot be handled with the usual techniques. In
Fig.~\ref{fig3}(a) we illustrate the phase-space volume of the
trapped region for $\varepsilon=0.0001$, indicating the location of
some $\varepsilon=0$ resonances as a guide, and in Fig.~\ref{fig3}(b)
we show a detail of the resulting ring. In comparison to
Fig.~\ref{fig1}(a), the gaps at the 1:3 and 1:6 SR are wider, but the
overall structure remains. The histogram is divided in three distinct
regions separated by those SR; $\varepsilon$ tunes the width of the
gaps. We construct the ring using different colors for the initial
conditions in the different $\langle \Delta t \rangle$ regions. The
ring displays two well--separated components. Each one is related to a
different region in phase space; this is a consequence of the large
gap opened by the 1:3 SR. These components are entangled and form a
braided ring~\cite{Merlo07}.

We note in Fig.~\ref{fig3}(b) that there are only two independent ring
components instead of three, as expected from the three disjoint
intervals in $\langle \Delta t\rangle$ defined by the 1:3 and 1:6
resonance. This follows from the fact that the 1:6 resonance does not
separate enough the phase-space regions around it, so the projections
into the $X-Y$ plane of these two intervals overlap and do not
manifest two components. Yet, as observed in Fig.~\ref{fig3}(b), the
corresponding ring component displays a clear separation of the ring
particles, except for a thin common strip, depending on which side of
the 1:6 resonance they come from. For larger values of $\varepsilon$
we have observed rings with three strands~\cite{Merlo07}. Therefore,
multiple ring components are obtained by exciting SR through nonzero
eccentricity. Note that these properties follow from the higher
dimensionality of phase space and nonlinear effects.

The fact that the overall structure of Fig.~\ref{fig3}(a) is similar
to the two DOF case and to the population histogram of asteroids may
be an indication of universality for higher dimensions. Yet, our
results are inconclusive for this issue: Fig.~\ref{fig3}(a) represents
the first results of the phase-space volume occupied by trapped orbits
for a system of more than two DOF.

To summarize, we have studied the phase-space volume occupied by
trapped orbits using a scattering billiard, a disk rotating on a
Kepler orbit, and succeeded in relating some of its structure to the
occurrence of SR. For two DOF the universal structure of the
phase-space volume occupied by trapped orbits is extended to
scattering systems~\cite{Contopoulos,Dvorak05}. For more DOF the
qualitative similarity of Fig.~\ref{fig3}(a) and Fig.~\ref{fig1}(a) is
the first indication that universality may hold also for higher
dimensions; yet, this issue remains open. Stability resonances are
defined by a resonant condition on the stability exponents of the
linearized dynamics around a central stable periodic orbit. These
resonances, and not the mean-motion resonances, manifest locally as a
reduction of the phase-space volume of the trapped trajectories due
to nonlinear effects. While SR have a local manifestation, they have
global consequences: Nonlinear and higher-dimensional effects
(nonvanishing eccentricity of the disk's motion) lead to an effective
separation of the trapping region in phase space
[Fig.~\ref{fig3}(a)]. In the context of planetary rings, if this
separation is large enough, this yields a ring with two or more
components or strands, which may entangle and form a braided
ring. This provides a simple explanation of recent observations of
planetary rings with multiple components~\cite{rings}. These results
should be interesting beyond the context of planetary rings, in
systems where resonances and the phase-space structure are
significant; examples include particle
accelerators~\cite{aceleradores} and galactic
dynamics~\cite{galaxias}.

\begin{acknowledgments}
  We thank to R.~Dvorak, T.H.~Seligman and C. Sim\'o for useful
  discussions, and the support by the projects IN-111607 (DGAPA) and
  43375 (CONACyT). O.~Merlo is a postdoctoral fellow of the Swiss
  National Foundation (PA002-113177).
\end{acknowledgments}

\end{document}